\documentclass[conference]{IEEEtran}
\IEEEoverridecommandlockouts
\usepackage{cite}
\usepackage{amsmath,amssymb,amsfonts}
\usepackage{algorithmic}
\usepackage{graphicx}
\usepackage{textcomp}
\usepackage{xcolor}
\usepackage{hyperref}
\def\BibTeX{{\rm B\kern-.05em{\sc i\kern-.025em b}\kern-.08em
    T\kern-.1667em\lower.7ex\hbox{E}\kern-.125emX}}
\begin{document}

\title{The Overview of Segmental Durations Modification Algorithms on Speech Signal Characteristics}

\author{
    \IEEEauthorblockN{Kyeomeun Jang, Jiaying Li, Yinuo Wang}
\thanks{All the authors are co-authors of this work with School of Electrical and Computer Engineering, Georgia Institute of Technology, Atlanta, GA 30332, USA.}.
}
\maketitle
\begin{abstract}
This paper deeply evaluates and analyzes several mainstream algorithms that can arbitrarily modify the duration of any portion of a given speech signal without changing the essential properties (e.g., pitch contour, power spectrum, etc.) of the original signal. Arbitrary modification in this context means that the duration of any region of the signal can be changed by specifying the starting and ending time for modification or the target duration of the specified interval, which can be either a fixed value of duration in the time domain or a scaling factor of the original duration. In addition, arbitrary modification also indicates any number of intervals can be modified at the same time.
\end{abstract}
\begin{IEEEkeywords}
Time-scale signal modification, arbitrary segmental speech modification, algorithm evaluation, speech signal processing
\end{IEEEkeywords}
\section{Introduction}
With a history of over 80 years, algorithms for time-scale modification (TSM) of speech have significant applications in speech data compression\cite{driedger2016review}, music signal playback and so on\cite{driedger2011time}. The most common studied methods are time domain modification, frequency domain modification, and model based modification.

Although there are lots of well-studied time-scale modification algorithms, the need for arbitrary modification is still not satisfied. Users always hope to modify their files wherever they want during speech editing and audio mixing. 

In this paper, we realized the algorithm which allows arbitrary modification of the duration of any portion of a given speech signal without changing the essential properties. With a well-designed graphical user interface, users can input a ``contour'' which defines the
durational expansion or contraction factor as a function of time and get their expected modified speech. After various evaluations, our program is highly compatible and robust with powerful functions.
\section{Theoretical Foundation}
To achieve the goal of modifying the speech duration, we need to modify the length of the speech signal without changing other speech characteristics such as speech contour and pitches. Generally, there are three main methods:
\begin{itemize}
    \item Time domain modification: OLA, SOLA, SOLAFS, TD-PSOLA, WSOLA and so on;
    \item Frequency domain modification: MSTFTM and so on;
    \item Model based modification: Phase Vocoder, Sinusoidal Modeling and so on.
\end{itemize}
The basic idea of OLA is to divide the speech into several consecutive non-overlapping frames, and then repeat or discard some of the speech frames to slow down or speed up the speech rate. OLA simply repeats or discards speech frames, which results in discontinuous waveform between two adjacent frames and leads to the broken pitch, resulting in poor speech quality. 

Synchronous Waveform Superposition Method (SOLA)\cite{makhoul1986time} is proposed to solve the discontinuity of the waveform. The algorithm is mainly divided into two stages: framing and synthesis. The framing stage completes the windowing task of the original speech signal. In order to reduce the discontinuity phenomenon, windowing and smoothing are generally performed at the same time . SOLAFS\cite{hejna1991solafs}, TD-PSOLA \cite{moulines1990pitch}, WSOLA\cite{verhelst1993overlap} and other SOLA algorithms are based on SOLA, and they all have some different improvements.

The most representative method in the frequency domain method is MSTFTM ( Modified Short-Time Fourier Transform Magnitude)\cite{griffin1984signal,roucos1985high}, which is based on the short-time Fourier transform. According to the principle of least mean square error, the short-time Fourier transform amplitude spectrum of a time-domain signal is found to approximate the spectrum of an ideal variable-speed signal.

The basic idea of model based method is to establish a model of the speech signal, and then modifying the relevant parameters of the model according to the needs to achieve the purpose of changing the pitch or duration of a speech. The phase vocoder method decomposes the speech into several sinusoidal signals through a band-pass filter, and then adjust the amplitude and phase of the sinusoidal signal to expand or compress the speech signal\cite{laroche1999new}. As for the sinusoidal modeling method, it is quite similar to the phase vocoder method. It needs to estimate the instantaneous amplitude, instantaneous frequency, instantaneous phase and other parameters of the model. However, the algorithm complexity is too high, especially comparing to its synthesis performance.

In this project, we will choose three typical algorithms for comparison: Phase vocoder, SOLAFS and WSOLA.
\subsection{Phase Vocoder\cite{laroche1999new,portnoff1976implementation,dolson1986phase,ellis2002phase}}
There are two stages for phase vocoder. In the first stage, prominent peaks are identified, and the frequency axis is divided into ”regions of influence” dominated by each peak. In the second stage, the regions around each peak are shifted, or translated, to new locations, thus achieving the desired frequency modification. 

Suppose that the frequency of the input speech signal $x(n)$ is $\omega$, then the expression of $x(n)$ can be defined as:
\begin{equation}
x(n)=Ae^{j\omega n+j\phi}\label{eq1}
\end{equation}
Denote $h(n)$ as the phase vocoder analysis window, and $H(\Omega)$ is the Fourier transform for $h(n)$ at frequency $\Omega$. The STFT of $x(n)$ at time $t^u_a$ and frequency $\Omega$ is:
\begin{equation}
X(\Omega,t^u_a)=AH(\Omega-\omega)e^{j\phi}\label{eq2}
\end{equation}
Suppose the speech signal around $\omega$ is shifted by $\Delta\omega$, then the short-term output signal at time $t^u_a$ and frequency $\Omega$ corresponding to $Y(\Omega, t^u_a)$ is:
\begin{equation}
y_u(n)=h(n)Ae^{j\phi}e^{j(\omega+\Delta\omega)n}\label{eq3}
\end{equation}
Suppose that there are $R$ samples between two frames, the hop size for the phase vocoder is $\Delta\omega R$.
Simply detect the pitch by searching the maximum magnitude in its left and right neighbors. For a target time scaling factor $\beta$ which gives the desired compression or expansion rate:
\begin{itemize}
    \item Large FFT sizes, low sampling rate: $\Delta\omega=\omega(\beta-1)$ ;
    \item For other cases: $\Delta\omega=\alpha\omega^2+\omega(\beta-1)$, where $\alpha$ is a quadratic frequency mapping coefficient.
\end{itemize}
 Once the $\Delta\omega$ is known, we can shift the peaks and adjust the phases. This can be accomplished by simply multiplying by a complex hop size. The rotations should be cumulated from one frame to the next in order to maintain phase-coherence from one frame to the next:
 \begin{equation}
Z_{u+1}=Z_u\Delta\omega_{u+1} R\label{eq4}
\end{equation}
where R is the phase-vocoder hop size.
\subsection{SOLA\cite{hejna1991solafs,roucos1985high,makhoul1986time}}
Denote the input speech signal as $x[n]$, and split it into overlapping windows $x_m[n]$ with a fixed length $W$, and separated by a fixed analysis distance $S_A$:
\begin{equation}
x_m[n]=
\begin{cases}
x[mS_A+n] & \text{for}\ n=0,...,W-1\\
0 & \text{otherwise}
\end{cases}
\label{eq5}
\end{equation}
For target compression or expansion rate $\beta$, $\beta>1$ represents expansion and $\beta<1$ represents compression. The scaling factor $\alpha=1/\beta$, has opposite meaning. The windows have a new synthesis distance $S_S$, which is defines as $S_S=\beta S_A$. The synthesis shift used for each window $x_m[n]$ is allowed to vary by an amount $k_m$ in order to maximize the similarity of the data in the overlapping regions before the overlap-add step. Suppose output modification frames:
\begin{equation}
y_m[n]=y[mS_S+k_m+n]\label{eq6}
\end{equation}
then $y_m[n]$ can be recursively defined by:
\begin{equation}
\begin{cases}
\beta_m[n]y_m[n]+(1-\beta_m[n])x_m[n]&  n=0,...,W^m_{OV}-1\\
x_m[n] & n=W^m_{OV},...W-1
\end{cases}
\label{eq7}
\end{equation}
where $W^m_{OV}=k_{m-1}-k_m+W-S_S$.
\subsection{SOLAFS\cite{hejna1991solafs}}
SOLAFS is an improvement of SOLA. During the analysis, the windwos are chosen as:
\begin{equation}
x_m[n]=
\begin{cases}
x[mS_A+k_m+n] & \text{for}\ n=0,...,W-1\\
0 & \text{otherwise}
\end{cases}
\label{eq8}
\end{equation}
and the output $y_m[n]=y[mS_S+n]$ is recursively formed by:
\begin{equation}
\begin{cases}
\beta[n]y_m[n]+(1-\beta[n])x_m[n]&  n=0,...,W_{OV}-1\\
x_m[n] & n=W_{OV},...W-1
\end{cases}
\label{eq9}
\end{equation}
where $W_{OV}=W-S_S$ is the number of points in the overlap region.

A common used similarity measure is the normalized cross-correlatino between x and y in the overlap region:
\begin{equation}
k_m\gets \max \limits_{0\leq k\leq K_{max}}R^m_{xy}[k]\label{eq10}
\end{equation}
where $K_{max}$ is the maximum allowable shift from the window's initial starting position, and
\begin{equation}
R^m_{xy}[k]=\frac{r^m_{xy}[k]}{\sqrt{r^m_{xy}[k]r^m_{yy}}}\label{eq11}
\end{equation}
where
\begin{equation}
r^m_{xy}[k]=\sum_{n=0}^{W_{OV}-1}x[mS_A+k+n]y[mS_S+n]\label{eq12}
\end{equation}
\begin{equation}
r^m_{xx}[k]=\sum_{n=0}^{W_{OV}-1}x^2[mS_A+k+n]\label{eq13}
\end{equation}
\begin{equation}
r^m_{yy}=\sum_{n=0}^{W_{OV}-1}y^2[mS_S+n]=x[mS_A+\tau_m+n]\label{eq14}
\end{equation}
where $\tau_m=k_{m-1}+S_S-S_A$. Suppose $0\leq\tau_m\leq K_{max}$, at the $m^{th}$ shift, $k_m$ should be determined by 
\begin{equation}
\begin{cases}
\tau_m=k_{m-1}+(S_S-S_A) & \text{if\ } 0\leq \tau_m\leq K_{max}\\
\max \limits_{0\leq k\leq K_{max}}R^m_{xy}[k] & \text{otherwise}
\end{cases}
\label{eq15}
\end{equation}
\subsection{WSOLA\cite{verhelst1993overlap,driedger2014tsm}}
Based on SOLA algorithm, the STFT of input speech signal $x(n)$ after adding a window becomes:
\begin{equation}
X(\omega,m)=\sum_{n=-\infty}^{+\infty}x(n+m)w(n)e^{-j\omega n}\label{eq16}
\end{equation}
WSOLA uses three measures for the best similar waveform searching algorithm:
\begin{itemize}
    \item Cross-correlation coefficient;
    \item Normalised cross-correlation coefficient;
    \item Cross AMDF coefficient.
\end{itemize}
Among the three algorithms, WSOLA has the theoretical highest performance.
\section{Algorithm Comparison Framework Design}
\begin{figure}[h]
\centerline{\includegraphics[scale=0.24]{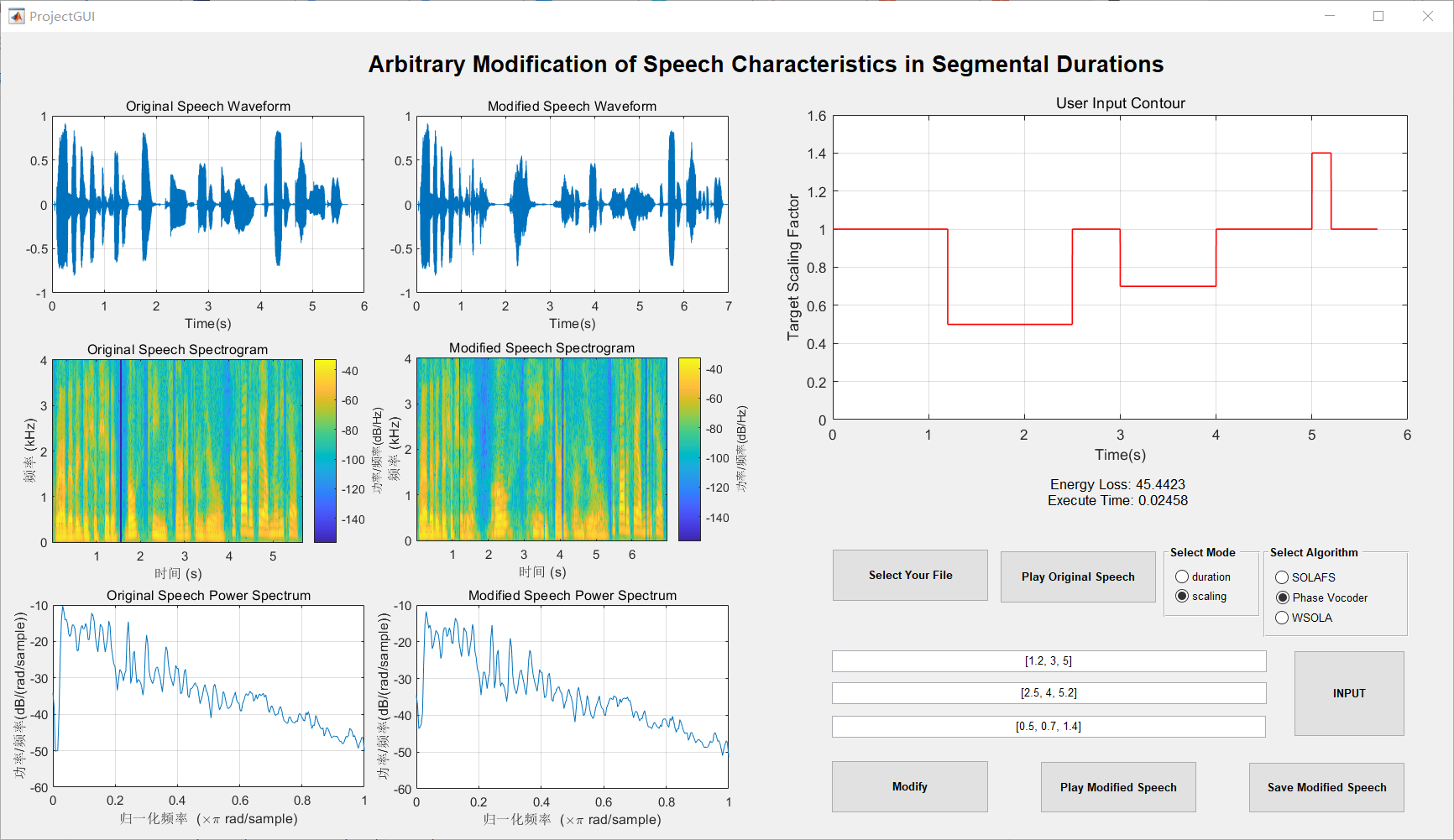}}
\caption{Graphic User Interface Panel}
\label{GUI}
\end{figure}

\begin{figure}[h]
\centerline{\includegraphics[scale=0.3]{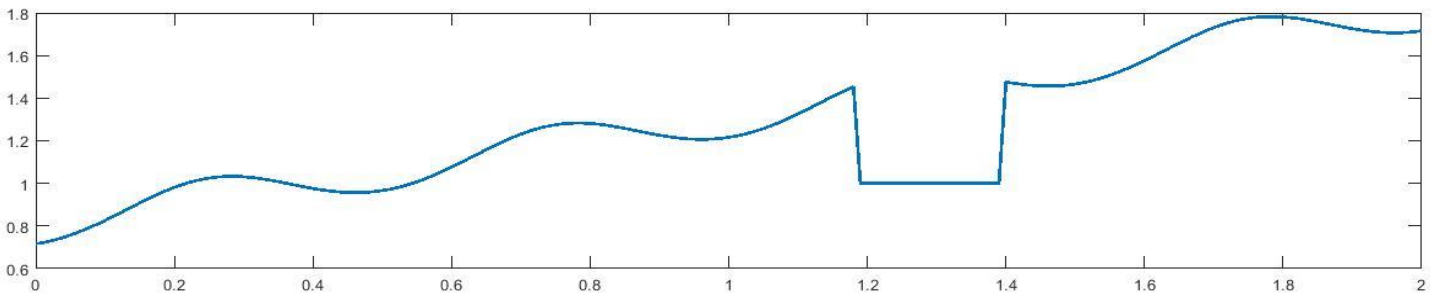}}
\caption{Contour Input Example. The x-axis indicates the time, and the y-axis represents the scaling factor }
\label{contour}
\end{figure}

To evaluate and compare above algorithms, we designed the framework to execute the signal modification based on specified start times, end times, target duration (scaling factors or desired duration), and algorithms (SOLAFS, PV-TSM, WSOLA) in the graphic interface panel looks like Fig.~\ref{GUI}.
A user input contour example is shown as Fig.~\ref{contour}. The input contour can be segmented and input as the duration arrays.

\begin{figure}[h]
\centerline{\includegraphics[scale=0.37]{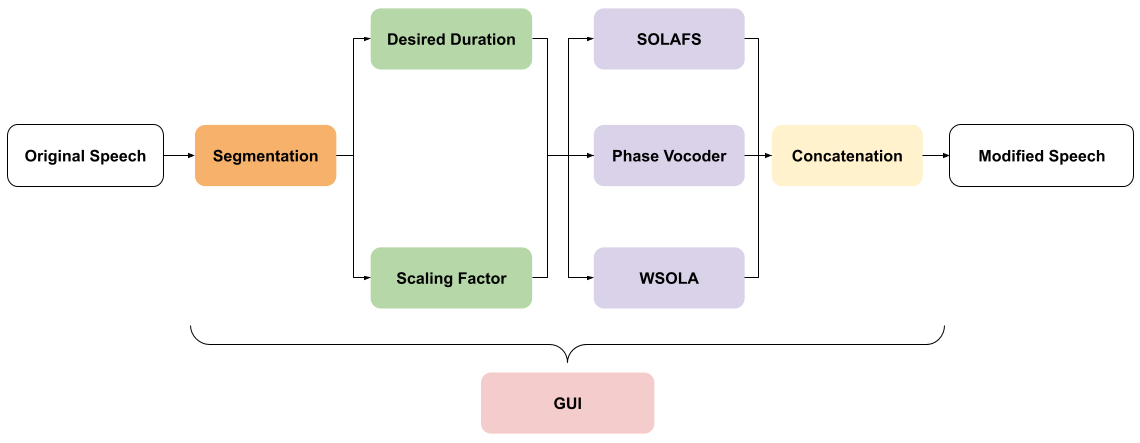}}
\caption{System Workflow}
\label{system}
\end{figure}

The main signal modification process is implemented in the seg\_modify function, and the system pipeline is shown in Fig.~\ref{system}. First, the digital speech signal is segmented into multiple arbitrary fragments based on the user-specified start times and end times and then the time domain input is transformed into discrete samples. For the target duration, our system allows two different input types. Specifically, in the duration mode, one can determine the desired duration of each selected segment defined by start time and end time, and in the scaling mode, user’s input of target duration represents the scaling factor of the original signal duration. But both two types of input will be unified as the scaling factor for the visualization as user input contour in the GUI and internal level algorithm implementation.

Then, each block of the segmented signal will be inserted into the modification algorithm that is being analyzed. After executing the time-scale modification, the modified segments are concatenated with the rest of the unselected original segments to build up the output speech signal.
\section{Result and Evaluation}
\subsection{Experiment Result}
\begin{figure}[h]
\centerline{\includegraphics[scale=0.30]{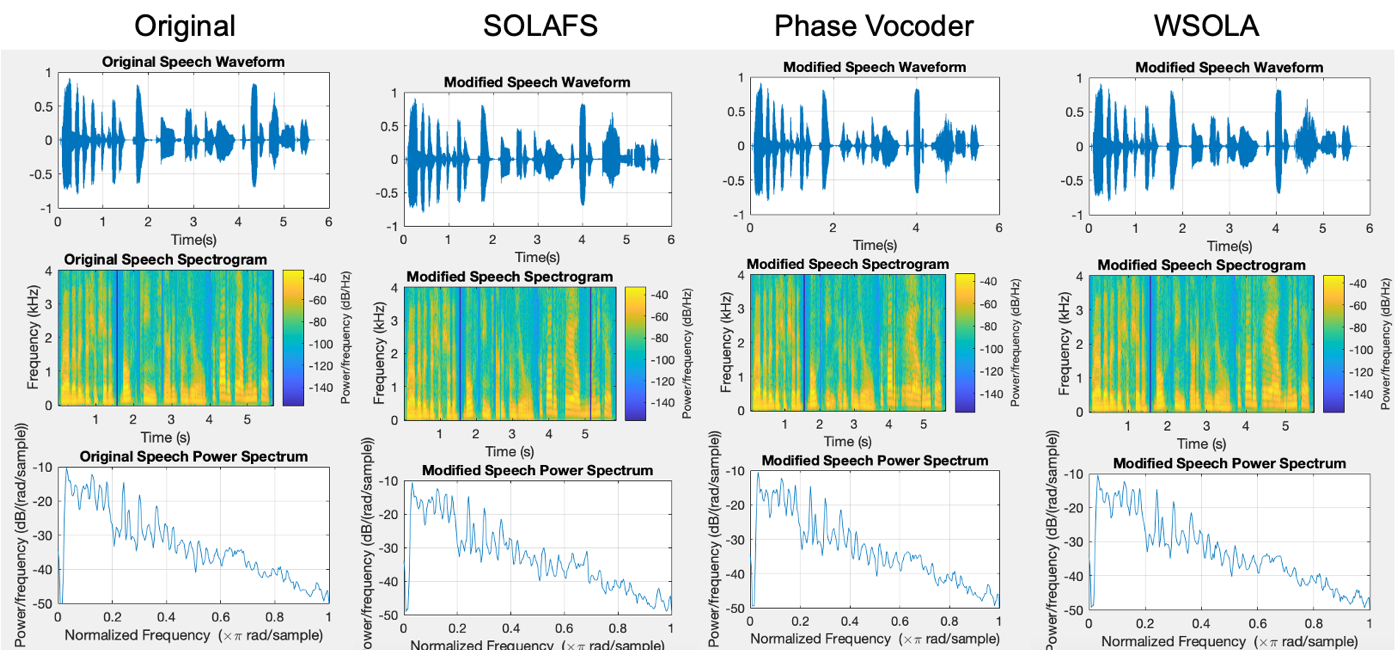}}
\caption{Modification Result of Different Algorithms}
\label{Result}
\end{figure}

All the above three algorithms are implemented and test with our framework, and Fig.~\ref{Result} shows an experiment result from one of our test cases where the scaling factor is 1.5 from 2-3 seconds, and 0.5 from 4.5-5 seconds. All of them could modify the speech segmental duration as well as keep the original speech characters. The waveform, spectrogram and power spectral density spectrum of original and modified speech signals are shown as below to demonstrate the performance of these three algorithms in time-scale speech signal modification.
\subsection{Performance Evaluation}
Although each of the above algorithms can satisfy the challenge requirements, there still are some differences among them, which need to be illustrated by using some quantitative comparisons. Specifically, we use three indicators, energy loss, execution time and difference of power spectral density, to evaluate their corresponding performance. 

Signal energy is a commonly used measurement in speech signal processing. Traditionally, the energy of a signal is defined using the square of the signal magnitude. Here, we define energy loss as the difference between the original speech and the modified speech signal:
\begin{equation}
E_{loss}=E_{in}-E_{out}=\sum_{n=-\infty}^{+\infty}|x_{i}(n)|^2-\sum_{n=-\infty}^{+\infty}|x_{o}(n)|^2\label{eq17}
\end{equation}
The power spectral density (PSD) provides an easier way of representing the distribution of signal frequency components than the DFT. We used Welch’s method\cite{welch1967use} to estimate the PSD of the speech signal and compute the difference between the original signal PSD and the modified signal PSD.

In addition, as a signal processing algorithm, execution efficiency is another crucial indicator we need to consider. Therefore, we implemented an experiment to compare the performance of different algorithms by using the above indicators and controlling all other variables the same, and the experiment results are shown in Table.~\ref{tab1}. 
\begin{table}[h]
\caption{Performance Comparison of Different Algorithms}
\begin{center}
\begin{tabular}{c c c c}
\hline\hline
{}& {SOLAFS} & {Phase Vocoder} & {WSOLA}\\\hline
{Energy Loss (dB)} & {-26.1494} & {55.8216} & {-1.5239}\\\hline
{Difference of PSD (W/Hz)} & {0.1220} & {0.1106} & {0.0743}\\\hline
{Execution Time (ms)} & {6.86} & {6.38} & {14.73}\\
\hline\hline
\end{tabular}
\label{tab1}
\end{center}
\end{table}
Summarized results are listed as follows:
\begin{itemize}
    \item WSOLA is the most robust model among the three algorithms. There was the smallest energy loss and power spectrum density change. Although it yields an accurate result, the execution time is longer than the other models depending on the delta values.
    \item SOLAFS has greater energy loss and difference in power spectrum density than WSOLA. However, it was more computationally efficient than WSOLA as the execution time was less than half of the WSOLA's.
    \item Phase Vocoder turns out to be the weakest model. The execution time and power spectrum density differences were similar to those of SOLAFS, but there was a large energy loss.
\end{itemize}
\section{Conclusion}
In this paper, we evaluated and implemented three mainstream arbitrary segmental duration modification algorithms of speech signals. After comprehensive comparison and assessments, we obtained the conclusion that among the three algorithms, WSOLA algorithm has the highest performance while SOLAFS and Phase Vocoder algorithm are more computationally efficient.

Due to the limited time and space, we did not involve more algorithms. In future work, it can be considered to add more algorithms into the existing framework and further analyze each of them.

\bibliographystyle{IEEEtran}
\bibliography{reference}{}
\end{document}